\documentclass[twoside,twocolumn,9pt]{article}
\usepackage{extsizes}
\usepackage[super,sort&compress,comma]{natbib} 
\usepackage[version=3]{mhchem}
\usepackage[left=1.5cm, right=1.5cm, top=1.785cm, bottom=2.0cm]{geometry}
\usepackage{balance}
\usepackage{mathptmx}
\usepackage{sectsty}
\usepackage{graphicx} 
\usepackage{lastpage}
\usepackage[format=plain,justification=justified,singlelinecheck=false,font={stretch=1.125,small,sf},labelfont=bf,labelsep=space]{caption}
\usepackage{float}
\usepackage{fancyhdr}
\usepackage{fnpos}
\usepackage[english]{babel}
\addto{\captionsenglish}{%
  
}
\usepackage{array}
\usepackage{droidsans}
\usepackage{charter}
\usepackage[T1]{fontenc}
\usepackage[usenames,dvipsnames]{xcolor}
\usepackage{setspace}
\usepackage[compact]{titlesec}
\usepackage{hyperref}
\usepackage{upgreek}

\definecolor{cream}{RGB}{222,217,201}

\begin{document}

\pagestyle{fancy}
\thispagestyle{plain}
\fancypagestyle{plain}{
\renewcommand{\headrulewidth}{0pt}
}

\makeFNbottom
\makeatletter
\renewcommand\LARGE{\@setfontsize\LARGE{15pt}{17}}
\renewcommand\Large{\@setfontsize\Large{12pt}{14}}
\renewcommand\large{\@setfontsize\large{10pt}{12}}
\renewcommand\footnotesize{\@setfontsize\footnotesize{7pt}{10}}
\makeatother

\renewcommand{\thefootnote}{\fnsymbol{footnote}}
\renewcommand\footnoterule{\vspace*{1pt}%
\color{cream}\hrule width 3.5in height 0.4pt \color{black}\vspace*{5pt}} 
\setcounter{secnumdepth}{5}

\makeatletter 
\renewcommand\@biblabel[1]{#1}            
\renewcommand\@makefntext[1]%
{\noindent\makebox[0pt][r]{\@thefnmark\,}#1}
\makeatother 
\renewcommand{\figurename}{\small{Fig.}~}
\sectionfont{\sffamily\Large}
\subsectionfont{\normalsize}
\subsubsectionfont{\bf}
\setstretch{1.125} 
\setlength{\skip\footins}{0.8cm}
\setlength{\footnotesep}{0.25cm}
\setlength{\jot}{10pt}
\titlespacing*{\section}{0pt}{4pt}{4pt}
\titlespacing*{\subsection}{0pt}{15pt}{1pt}

\fancyfoot{}
\fancyfoot[LO,RE]{\vspace{-7.1pt}\includegraphics[height=9pt]{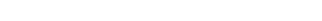}}
\fancyfoot[RO]{\footnotesize{\sffamily{ ~\textbar  \hspace{2pt}\thepage}}}
\fancyfoot[LE]{\footnotesize{\sffamily{\thepage~\textbar\hspace{4.65cm} }}}
\fancyhead{}
\renewcommand{\headrulewidth}{0pt} 
\renewcommand{\footrulewidth}{0pt}
\setlength{\arrayrulewidth}{1pt}
\setlength{\columnsep}{6.5mm}
\setlength\bibsep{1pt}

\makeatletter 
\newlength{\figrulesep} 
\setlength{\figrulesep}{0.5\textfloatsep} 

\newcommand{\topfigrule}{\vspace*{-1pt}%
\noindent{\color{cream}\rule[-\figrulesep]{\columnwidth}{1.5pt}} }

\newcommand{\botfigrule}{\vspace*{-2pt}%
\noindent{\color{cream}\rule[\figrulesep]{\columnwidth}{1.5pt}} }

\newcommand{\dblfigrule}{\vspace*{-1pt}%
\noindent{\color{cream}\rule[-\figrulesep]{\textwidth}{1.5pt}} }

\makeatother

\twocolumn[
  \begin{@twocolumnfalse}
\vspace{1em}
\sffamily
\begin{tabular}{m{4.5cm} p{13.5cm} }

& \noindent\LARGE{\textbf{A novel method to determine the phase-space distribution of a pulsed molecular beam}} \\
\vspace{0.3cm} & \vspace{0.3cm} \\

 & \noindent\large{Maarten~C.~Mooij,\textit{$^{a,b}$} Hendrick~L.~Bethlem,$^{\ast}$\textit{$^{a,c}$} Alexander~Boeschoten,\textit{$^{b,c}$} Anastasia~Borschevsky,\textit{$^{a,c}$} Ties~H.~Fikkers,\textit{$^{b,c}$} Steven~Hoekstra,\textit{$^{b,c}$} Joost~W.~F.~van~Hofslot,\textit{$^{b,c}$} Klaus~Jungmann,\textit{$^{b,c}$} Virginia~R.~Marshall,\textit{$^{b,c}$} Thomas~B.~Meijknecht,\textit{$^{b,c}$} Rob~G.~E.~Timmermans,\textit{$^{b,c}$} Anno~Touwen,\textit{$^{b,c}$} Wim~Ubachs,\textit{$^{a}$} and Lorenz~Willmann.\textit{$^{b,c}$} (NL-\textit{e}EDM collaboration) } \\

& \noindent\normalsize{We demonstrate a novel method to determine the longitudinal phase-space distribution of a cryogenic buffer gas beam of barium-fluoride molecules based on a two-step laser excitation scheme. The spatial resolution is achieved by a transversely aligned laser beam that drives molecules from the ground state $X^2\Sigma^+$ to the $A^2\Pi_{1/2}$ state around 860\;nm, while the velocity resolution is obtained by a laser beam that is aligned counter-propagating with respect to the molecular beam and that drives the Doppler shifted $A^2\Pi_{1/2}$ to $D^2\Sigma^+$ transition around 797\;nm. Molecules in the $D$-state are detected virtually background-free by recording the fluorescence from the $D-X$ transition at 413 nm. As molecules in the ground state do not absorb light at 797\;nm, problems due to due to optical pumping are avoided. Furthermore, as the first step uses a narrow transition, this method can also be applied to molecules with hyperfine structure. The measured phase-space distributions, reconstructed at the source exit, show that the average velocity and velocity spread vary significantly over the duration of the molecular beam pulse. Our method gives valuable insight into the dynamics in the source and helps to reduce the velocity and increase the intensity of cryogenic buffer gas beams. In addition, transition frequencies are reported for the $X-A$ and $X-D$ transitions in barium fluoride with an absolute accuracy below 0.3\;MHz. 
} \\

\end{tabular}

 \end{@twocolumnfalse} \vspace{0.6cm}

  ]

\renewcommand*\rmdefault{bch}\normalfont\upshape
\rmfamily
\section*{}
\vspace{-1cm}


\footnotetext[1]{\textit{$^{a}$~Department of Physics and Astronomy, LaserLaB, Vrije Universiteit Amsterdam, de Boelelaan 1081, 1081 HV Amsterdam, The Netherlands; E-mail: H.L.Bethlem@vu.nl}}
\footnotetext{\textit{$^{b}$~Nikhef, National Institute for Subatomic Physics, 1098 XG Amsterdam, The Netherlands}}
\footnotetext{\textit{$^{c}$~Van Swinderen Institute for Particle Physics and Gravity, University of Groningen, 9747 AG Groningen, The Netherlands}}




\section{Introduction}
Cold molecules offer unique possibilities for precision tests of fundamental physics theories~\cite{safronova2018a,andreev2018,roussy2023}
, quantum technology~\cite{demille2002,ni2018,blackmore2019, yu2019}, and studies of quantum effects in molecular collisions~\cite{brouard2014,tang2023}. Successful methods to create cold molecules include the deceleration of molecular beams using electric fields via the Stark effect~\cite{bethlem1999,vandemeerakker2012}, magnetic fields via the Zeeman effect~\cite{vanhaecke2007,narevicius2008,vandemeerakker2012}, and laser cooling using near resonant light~\cite{shuman2010, mccarron2018, fitch2021}. The phase-space density -- the number of molecules per position and velocity interval -- of decelerated molecular beams is proportional to, and for Stark and Zeeman deceleration limited by, the phase-space density of the initial beam~\cite{vandemeerakker2012}. Therefore, the success of deceleration techniques depends crucially on the brightness of the initial beam. Furthermore, as there is a compromise between the number of molecules that are decelerated and the deceleration rate, it is highly desirable that the beam has a low initial velocity. 
An effective way to create intense, slow beams of molecules and molecular radicals is the so-called cryogenic buffer gas beam source, first introduced by Maxwell \emph{et al.}~\cite{maxwell2005} and further developed by Patterson \emph{et al.}~\cite{patterson2007}, van Buuren \emph{et al.}~\cite{vanbuuren2009}, Hutzler \emph{et al.}~\cite{hutzler2011}, Truppe \emph{et al.}~\cite{truppe2018} and others. Over the last few years, we have constructed a cryogenic buffer gas cooled beam source that provides barium monofluoride (BaF) molecules for an experiment that will search for the electron's electric dipole moment (\emph{e}EDM)~\cite{aggarwal2018}. In the experiment, BaF molecules are decelerated using a 4.5\;m long travelling-wave Stark decelerator~\cite{aggarwal2021a}. In order to decelerate a reasonable fraction of the molecular beam, the velocity spread of the initial beam should be small, while the averaged forward velocity should be below $\sim$200\;m/s\cite{aggarwal2018}. Therefore, accurate knowledge of the velocity distribution is crucial while optimising the intensity of the source.   

A simple and general way to obtain the longitudinal velocity distribution is by measuring the time of flight profile of the molecular beam at two positions along the molecular beam path. The average velocity of the beam is determined from the time difference in the mean of these distributions, while the velocity spread is determined from the difference of their widths. This method works well if the molecular beam pulse is short compared to the flight path or if the velocity of the molecules is independent of the time they exit the source and accurately described by a Gaussian distribution~\cite{wang1988}. 

A more direct method to measure the velocity distribution is by measuring the Doppler shift of a transition using a laser beam that is counter-propagating with respect to the molecular beam. A downside of this method is that molecules will already be excited and may decay into dark states before the detection zone is reached. This effect will limit the signal strength that is obtained. Furthermore, as slow molecules are more likely to be pumped away, the measurement may not accurately reflect the true velocity distribution~\cite{bumby2017}. Optical pumping can be reduced by introducing a small angle between the molecular beam and the laser~\cite{barry2011}, which however, complicates the interpretation of the measurement as in this situation the detection volume will not be well defined. A more serious problem arises when the hyperfine splitting of the observed transition is comparable to the Doppler profile, resulting in different velocity components being excited at the same frequency of the light. In such a case, the integrated velocity distribution of the beam can still be obtained by a deconvolution procedure, but it is not possible to measure the velocity distribution at a specific time in the pulse. 


In this paper, we present a method to accurately measure the phase space distribution of a molecular beam using a two-step laser excitation scheme. In the first step, molecules are brought to an excited state, independent of their forward velocity, at a well-defined position along the molecular beam path using a laser beam that is perpendicular to the molecular beam. In the second step, the longitudinal velocity of molecules in this excited state is measured using a laser beam that is counter-propagating with respect to the molecular beam. This scheme avoids the problems mentioned above and allows us to record the complete phase space distribution of the molecular beam. We demonstrate this method for a cryogenic buffer gas cooled beam source of BaF molecules but it can be easily adapted to other molecules and other sources.

\section{Method}\label{sec:method}

\begin{figure}[t]
    \centering
    \includegraphics[width=\columnwidth]{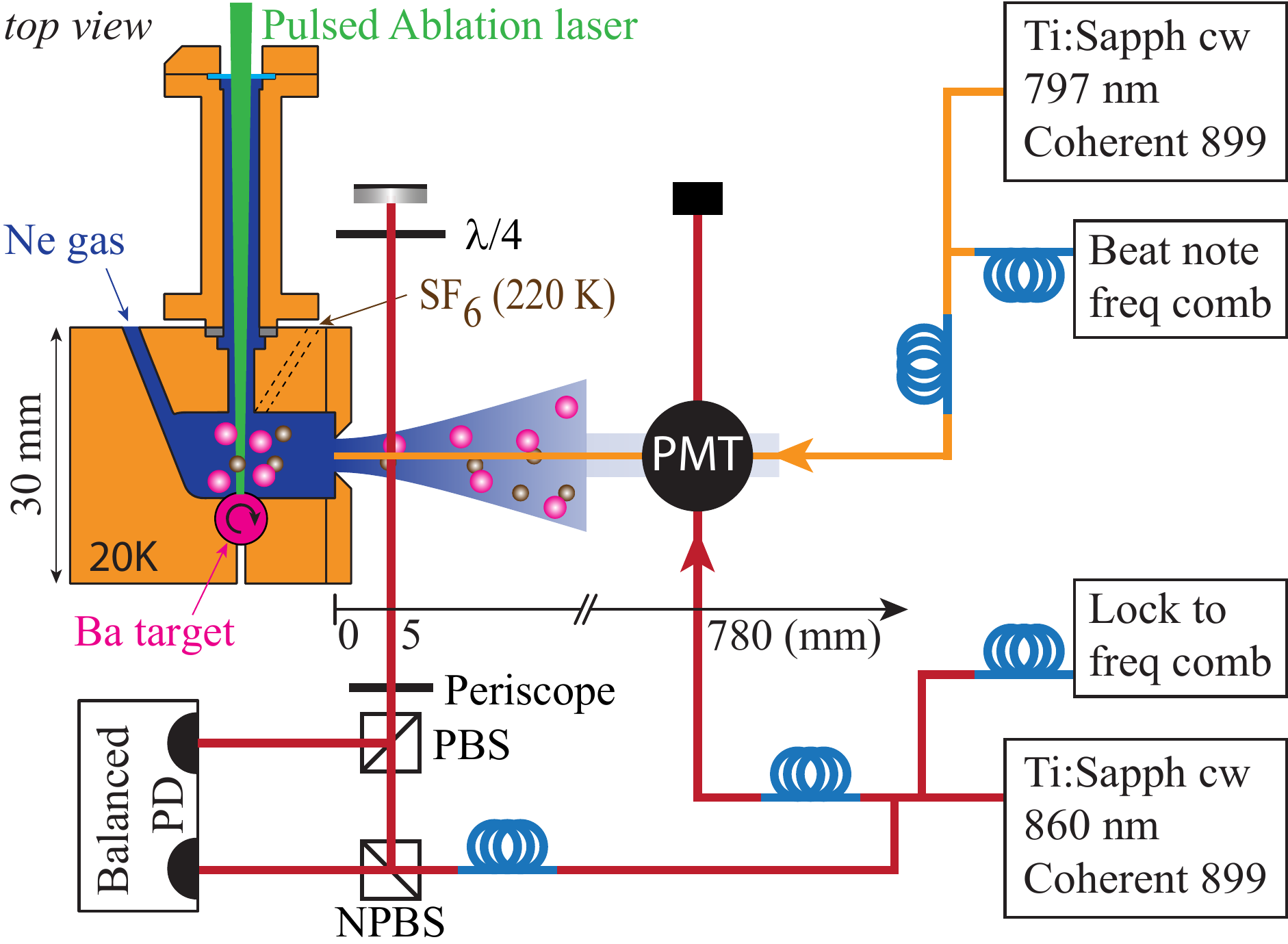}
    \caption{Schematic view of the experimental setup showing the cryogenic buffer gas beam source and the lasers used for absorption and fluorescence detection. Barium monofluoride molecules are created inside the cell and expand into the vacuum to form a molecular beam. The performance of the source is monitored via absorption detection 5\;mm behind the cell. The phase-space distribution of the beam is recorded 780\;mm after the cell using a two-step laser excitation scheme. In the first step, BaF is brought to an excited state, independently of their forward velocity, but at a well-defined position along the molecular beam path using a laser beam that is perpendicular to the molecular beam. In the second step, the longitudinal velocity of molecules in this excited state is measured using a laser beam that is counter-propagating with respect to the molecular beam. Fluorescence back to the ground state is measured using a photomultiplier tube (PMT).}
    \label{fig:setupVelocity}
\end{figure}

Figure~\ref{fig:setupVelocity} shows a schematic of our method including the top-view of our cryogenic buffer gas beam source and the optical beam paths. The design of our cryogenic source is based on that of Truppe et al.~\cite{truppe2018}. The heart of our setup is formed by a cubical copper cell kept at a temperature of around 20\;K using a 2-stage cryo-cooler (Sumitomo Heavy Industries, cold head RP-082B2S). A continuous flow of pre-cooled neon is passed through the cell at a flow rate of 20\;standard cubic centimeter per minute (sccm)~\footnote[2]{1\;\textsc{sccm} = 4.48$\times10^{17}$\;particles/s}. Within the cell, barium atoms are ablated by a pulsed Nd:YAG laser (532\;nm, 5\;ns pulse, 10\;Hz, 8\;mJ per pulse) from a rotating solid Ba target. The barium atoms in the plasma plume react with sulphur hexafluoride (SF$_{6}$) molecules, that are injected into the cell with a flow rate of typically 0.03\;\textsc{sccm} from a copper tube that is kept at a temperature of 220\;K. The BaF molecules created in this reaction are cooled via collisions with the neon atoms and form a molecular beam by expanding through a 4.5\;mm diameter orifice into vacuum. The cell is surrounded by a copper and an aluminium shield, at temperatures of 6\;K and $\sim$30\;K, respectively. Besides acting as heat shields, these cylinders provide the necessary pumping capacity to allow pressures on the order of 10$^{-2}$\;mbar inside the cell, while maintaining a pressure below 10$^{-6}$\;mbar in the molecular beam chamber.    

\begin{figure}[t]
    \centering
    \includegraphics[width=\columnwidth]{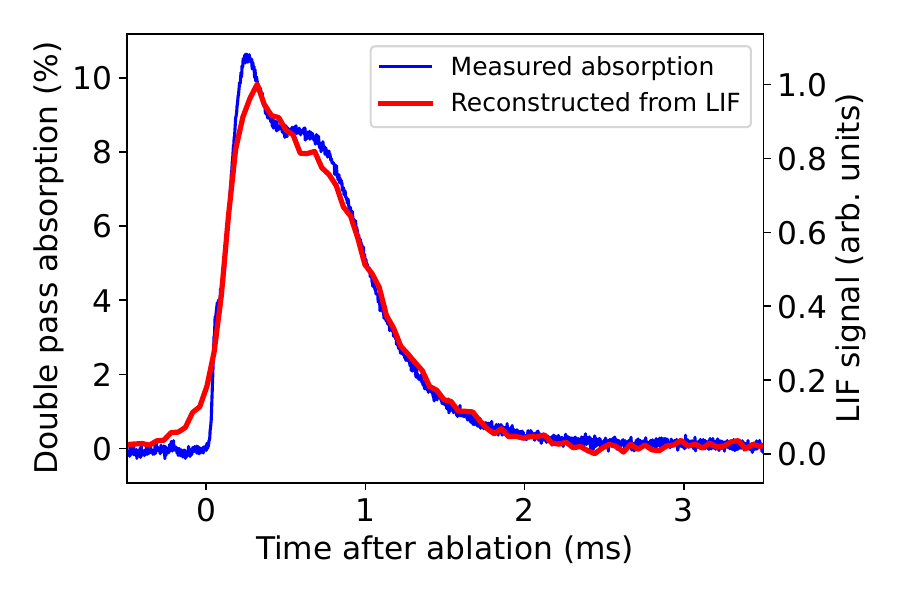}
    \caption{Simultaneously measured absorption and reconstructed fluorescence data as a function of time after ablation. The blue line presents the double-pass laser absorption signal measured 5\;mm behind the exit of the source. The red line corresponds to the time-of-flight of the molecules that is reconstructed from the phase-space distribution measurement further downstream the molecular beam path, with the vertical scale adjusted to match the absorption measurement.
    }
    \label{fig:absReconstucted}
\end{figure}

To monitor the performance of the source, the BaF molecules are detected using absorption 5\;mm behind the cell on the $X^2\Sigma^+,N=0,J=1/2 \rightarrow A^2\Pi_{1/2},J=1/2$ transition using a laser beam with a full with at half maximum (FWHM) diameter of 1\;mm 
and a power of $\sim$1\;$\upmu$W near 860\;nm from a Ti:Sapphire laser. Fig.~\ref{fig:absReconstucted} shows a typical absorption measurement (blue curve) together with a time-of-flight measurement that is reconstructed from laser-induced fluorescence measurements recorded simultaneously (red curve), which will be discussed later. The absorption signal can be converted into an absolute number by taking into account the spatial and velocity distributions of the beam in the longitudinal and transverse directions, using a procedure that is similar to the one described by Wright \emph{et al.}~\cite{wright2022}. The duration of the molecular beam pulse is about 1\;ms and the peak absorption is 10\% (double pass), which corresponds to $1.9(6)\times10^{10}$\;BaF molecules in the $N=0$ state per pulse and $1.3(5)\times 10^{11}$\;molecules per sr per pulse.

At a distance of 780\;mm from the source, in a second, differentially pumped, vacuum chamber, the molecules are excited by light from two Ti:Sapphire lasers (Coherent 899) that are referenced to a frequency comb (Menlo Systems FC1500-250-WG) that is stabilized to a caesium clock (Microsemi CSIII Model 4301B). The beat note of the lasers with the frequency comb is recorded every molecular shot with a Siglent SSA3021X spectrum analyser with an accuracy of typically 1.1\;MHz. 
One of the lasers is aligned perpendicular to the molecular beam and is resonant with the $X^2\Sigma^+ \rightarrow A^2\Pi_{1/2}$ transition around 860\;nm, while the other laser is aligned to be counter-propagating with respect to the molecular beam and resonant with the $A^2\Pi_{1/2} \rightarrow D^2\Sigma^+$ transition around 797\;nm. The frequency of the second laser is red-shifted with respect to the transition frequency to compensate for the Doppler shift. From this detuning, we can infer the longitudinal velocity of the molecules. Note that, rather than two subsequent one-photon transitions, the excitation process can be described as a two-photon transition from the $X$ to $D$-state that is enhanced by the intermediate $A$-state. Any detuning of the laser that is used to drive the $X-A$ transition, can be compensated by an equal but opposite detuning of the laser that is used to drive the $A-D$ transition. Therefore, frequency stabilization of both lasers is equally important.

\begin{figure}[t]
    \centering
    \includegraphics[width=0.8\columnwidth]{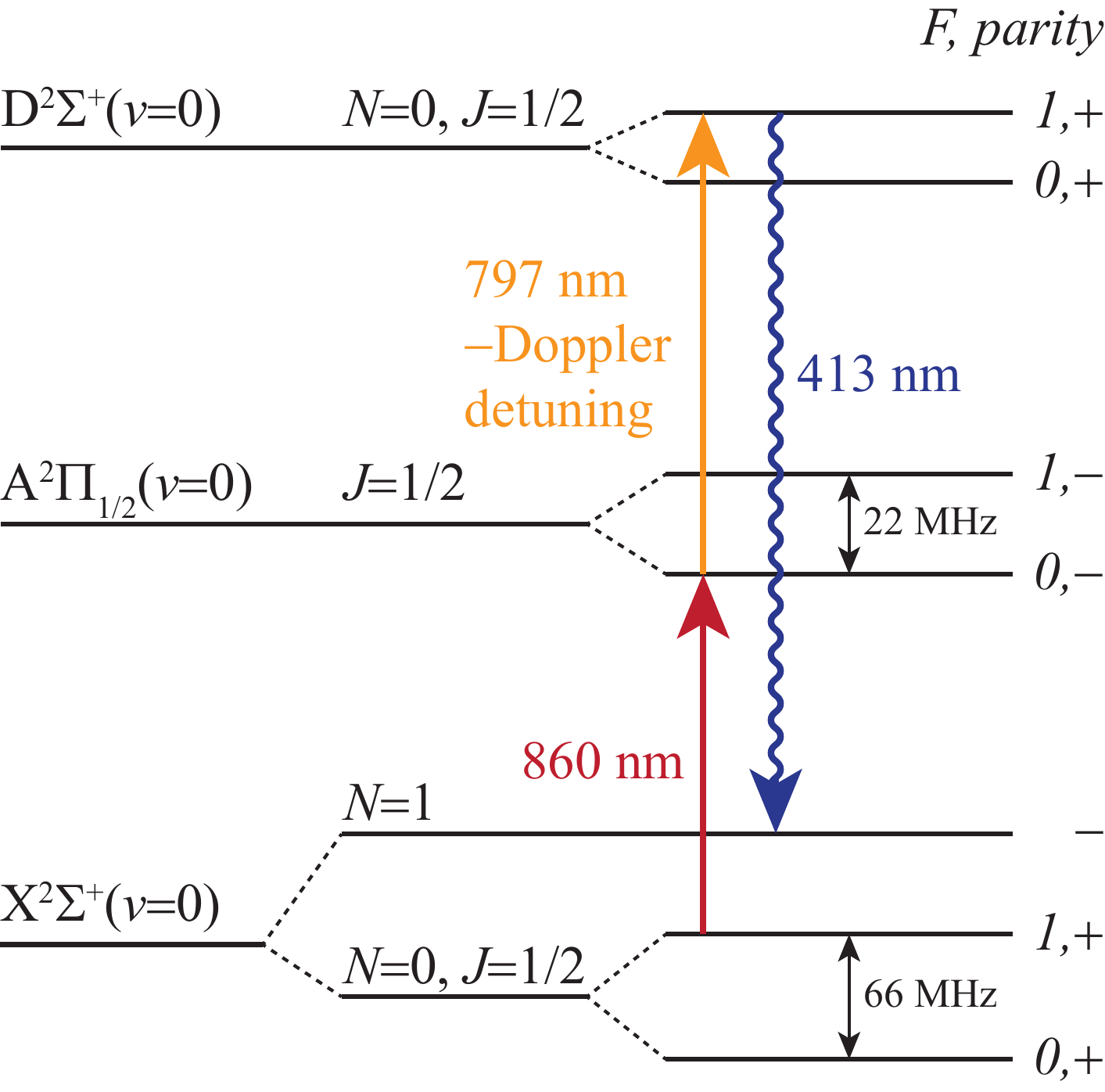}
    \caption{Energy level scheme of $^{138}$BaF showing the relevant energy levels. Barium fluoride molecules are exited to the $D$-state using two lasers around 860\;nm and 797\;nm. Molecules are detected using the fluorescence from the $D$-state back to the $X$-state at 413\;nm.}
    \label{fig:energyDiagram}
\end{figure}
  
Figure~\ref{fig:energyDiagram} shows the energy levels of $^{138}$BaF relevant for our experiment. The lowest rotational level of the $X^2\Sigma^{+}$ manifold, the $N=0,J=1/2$ state, is split into two hyperfine components, $F=0$ and $F=1$, due to the nuclear spin of the fluorine atom. The two components are separated by 65.8\;MHz~\cite{ernst1986}. We drive a transition from the $X^2\Sigma^+,N=0,J=1/2, F=1$ state to the $A^2\Pi_{1/2},J=1/2, F=0$ state using light around 860\;nm. In the $A$-state, the $F=0$ and $F=1$ levels are separated by 21.87\;MHz~\cite{denis2022}, sufficiently large to selectively drive a transition to the $F=0$ level using a laser beam that is perpendicular to the molecular beam. From the $A^2\Pi_{1/2},J=1/2, F=0$, we drive a transition to the $D^2\Sigma^+,N=0,J=1/2,F=1$ around 797\;nm using a second laser that is aligned to be counter-propagating with respect to the molecular beam. Note that, while the hyperfine splitting in the $D$-state is small, we can be sure to drive a single transition as the $A^2\Pi_{1/2},J=1/2,F=0 \rightarrow D^2\Sigma^+,N=0,J=1/2,F=0$ transition is not allowed\footnote[3]{We have explicitly looked for hyperfine structure in the $D$ state by using the $A^2\Pi_{1/2},J=1/2,F=1$ as intermediate state but were unable to resolve it within our experimental linewidth of 5\;MHz (FWHM). The mean transition frequency from $X$ via $A,F=0$ to only $D, F=1$ agrees within the uncertainty of 170\;kHz to the mean frequency when measuring a transition from $X$ via $A,F=1$ to both $D, F=0,1$.}.

Once excited to the $D$-state, part of the molecules will decay back to the ground state by emitting a photon at 413\;nm which is efficiently detected using a photomultiplier tube (PMT) (Thorn EMI 9558 QB). A bandpass filter around 400\;nm (Thorlabs FBH400-40) is used to filter out scattered photons from the laser beams and unwanted fluorescence, resulting in a nearly background-free detection~\cite{murphree2009}.

The laser used for driving the $X-A$ transition has a power of typically 0.1\;mW in a beam with a FWHM diameter of 2.2\;mm, corresponding to a peak intensity of $\sim2$\;mW/cm$^2$. 
This intensity is sufficient to obtain a good signal-to-noise ratio, while avoiding off-resonant excitation. The laser used for driving the $A-D$ transition has a power of about 10\;mW in a beam with a FWHM diameter of 2.2\;mm, corresponding to a peak intensity of $\sim$2$\times10^2$\;mW/cm$^2$. This intensity is sufficient to obtain a good signal-to-noise ratio, while avoiding power broadening which would lead to a decrease of the velocity resolution. Note that the intensity of the laser used to drive the $A-D$ transition is 100 times larger than that used to drive the $X-A$ transition as the excitation process needs to compete with the rapid decay of the $A$-state (which has a lifetime of 57.1\;ns~\cite{aggarwal2019}) back to the ground state.

Essential for our method is that molecules are only excited to the $D$-state at the exact location where the two lasers overlap; the laser that addresses the $X-A$ transition, which is aligned perpendicular to the molecular beam, determines the spatial resolution along the molecular beam axis, while the laser that addresses the $A-D$ transition, which is aligned counter-propagating with respect to the molecular beam, determines the velocity resolution. In this way, problems with optical pumping are avoided.

\section{Doppler free transition frequencies}\label{sec:Dopplerfree}

It is important to know the exact (Doppler-free) transition frequency of the $X-D$ transition accurately, as we relate the measured Doppler shifted frequency to the velocity. We have also measured the Doppler-free transition frequency of the $X-A$ transition to ensure that the 860\;nm laser is aligned perfectly perpendicular to the molecular beam axis.

\begin{figure}[t]
    \centering
    \includegraphics[width=\columnwidth]{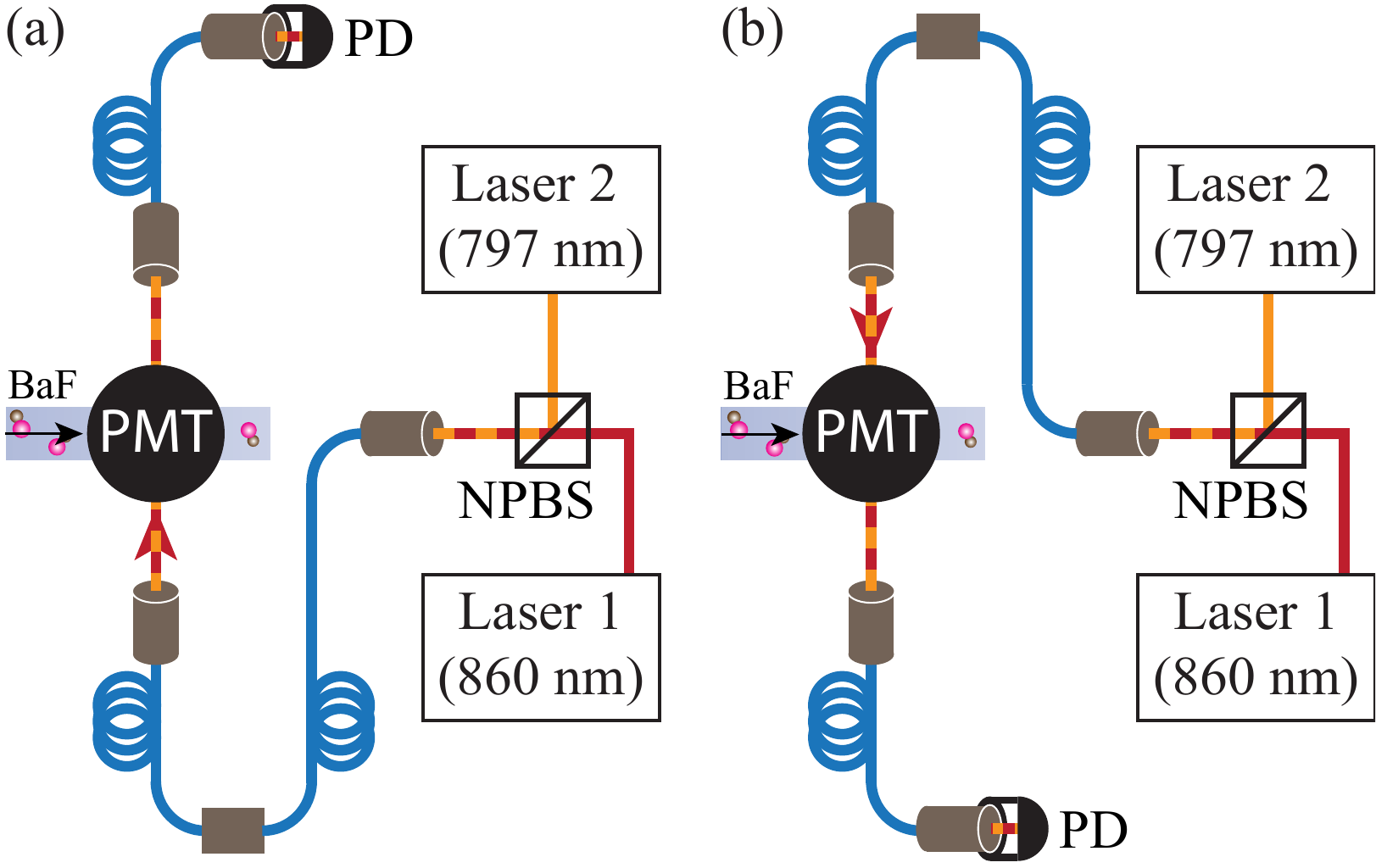}
    \caption{Experimental scheme for overlapping the path of two counter-propagating laser beams, which is used for measuring Doppler-free transition frequencies in a molecular beam. Two lasers are combined using a non-polarizing beam splitter (NPBS) and are coupled into an optical fiber. Using a mating sleeve, this fiber is connected to a second fiber that brings the light to the right-hand side (a) or the left-hand side (b) of the molecular beam setup. After crossing the molecular beam, the laser beams are coupled into a third optical fiber which is connected to a photodiode. By exchanging the first fiber with the photodiode, we can reverse the direction of the light, and hence the sign of the residual Doppler shift. 
    }
    \label{fig:setupDopplerFree}
\end{figure}

Figure~\ref{fig:setupDopplerFree} shows the setup used to determine the Doppler-free transition frequencies. Using a non-polarizing beam splitter (NPBS), the two lasers used for driving the $X-A$ and $A-D$ transitions are coupled into a single optical fiber. Using a mating sleeve, this fiber is connected to a second fiber that brings the light to the right-hand side (shown in Fig.~\ref{fig:setupDopplerFree}(a)) or the left-hand side (shown in Fig.~\ref{fig:setupDopplerFree}(b)) of the molecular beam setup. The laser beams are subsequently coupled out of the fiber and aligned such that they cross the molecular beam below the PMT. After crossing the molecular beam, the laser beams are coupled into a third optical fiber which is connected to a photodiode. Typically, 10\;$\%$ of the light that exits the second fiber is collected back into the third fiber. By exchanging the first fiber with the photodiode, we can reverse the direction of the light, i.e., we can switch between the situation depicted in Fig.~\ref{fig:setupDopplerFree}(a) and Fig.~\ref{fig:setupDopplerFree}(b). This procedure ensures that the path followed by the lasers in Fig.~\ref{fig:setupDopplerFree}(a) and Fig.~\ref{fig:setupDopplerFree}(b) exactly overlap. We estimate that the angle between the two laser beams coming from opposite directions is below 5$\times10^{-3}$\;degrees, which for a 200\;m/s beam corresponds to an uncertainty on the Doppler-free $X-A$ transition frequency of below 10\;kHz. 
Our method is a simplified version of the one recently employed by Wen \emph{et al.}~\cite{wen2023} to perform spectroscopy in a beam of metastable helium.  

The fact that the beam paths overlap does not imply that the lasers are exactly perpendicular to the molecular beam, i.e. the $X-D$ transition frequency measured with the laser beam coming from the right side might be Doppler shifted. However, this Doppler shift will be opposite to the one measured with the laser beam coming from the left side. The Doppler-free $X-D$ transition frequency is found by taking the average of the two measurements. The spectra were fitted to a Lorentzian function with a width of $\sim$5\;MHz resulting from the transverse velocity spread of the beam and the finite lifetime of the $D$-state. We have performed several of these measurements, using both vertical and horizontal polarization, over multiple days. The error is conservatively taken to be 0.3\;MHz, the maximum deviation of the weighted mean. This error is mainly limited by the lack of control over magnetic and electric fields in the interaction zone. The Doppler-free $X-A$ transition was obtained by following the same procedure, but by leaving out the filter in front of the PMT and having the 797\;nm laser blocked. 

\begin{table*}
    \centering
    \caption{Measured transition frequencies of the $X-D$ and $X-A$ transitions with standard deviation. The $A-D$ transition frequency is found by taking the difference between the $X-D$ and $X-A$
    } 
    \label{tab:energies}
    \begin{tabular}{lcll}
    \hline
    Transition & & & \;\;\;\;Frequency (MHz) \\
    \hline
    $X^2\Sigma^+,N=0,J=1/2,F=1$ & $\rightarrow$ & $D^2\Sigma^+,N=0,J=1/2,F=1$ &  \;\;\;\;724795734.10(14) \\
    $X^2\Sigma^+,N=0,J=1/2,F=1$ & $\rightarrow$ & $A^2\Pi_{1/2},J=1/2,F=0$ &  \;\;\;\;348666402.6(3) \\
    $A^2\Pi_{1/2},J=1/2,F=0$ & $\rightarrow$  & $D^2\Sigma^+,N=0,J=1/2,F=1$ &  \;\;\;\;376129331.5(3) \\
    \hline
    \end{tabular}
\end{table*}

The obtained transition frequencies are listed in Table~\ref{tab:energies} with the standard deviation quoted between brackets. Our value for the $X^2\Sigma^+,N=0,J=1/2,F=1 \rightarrow D^2\Sigma^+,N=0,J=1/2,F=1$ transition may be compared with the term energy determined by Effantin \emph{et al.}~\cite{effantin1990} from fluorescence spectra of highly excited rotational levels in the $D$-state. Our value is two orders in magnitude more accurate and deviates by $\sim$700\;MHz from the value found by Effantin \emph{et al.}. Our value for the $X^2\Sigma^+,N=0,J=1/2,F=1 \rightarrow A^2\Pi_{1/2},J=1/2,F=0$ transition is in good agreement with, but two orders in magnitude more accurate than the one found by Steimle \emph{et al.}~\cite{steimle2011} and Rockenh\"{a}user \emph{et al.}~\cite{rockenhauser2023}. The $A^2\Pi_{1/2},J=1/2,F=0 \rightarrow D^2\Sigma^+,N=0,J=1/2,F=1$ transition is found by subtracting the second and first entry of Table~\ref{tab:energies}.  

\section{Doppler shifted transition frequencies}\label{sec:Dopplershifted}

\begin{figure}[t]
    \centering
    \includegraphics[width=\columnwidth]{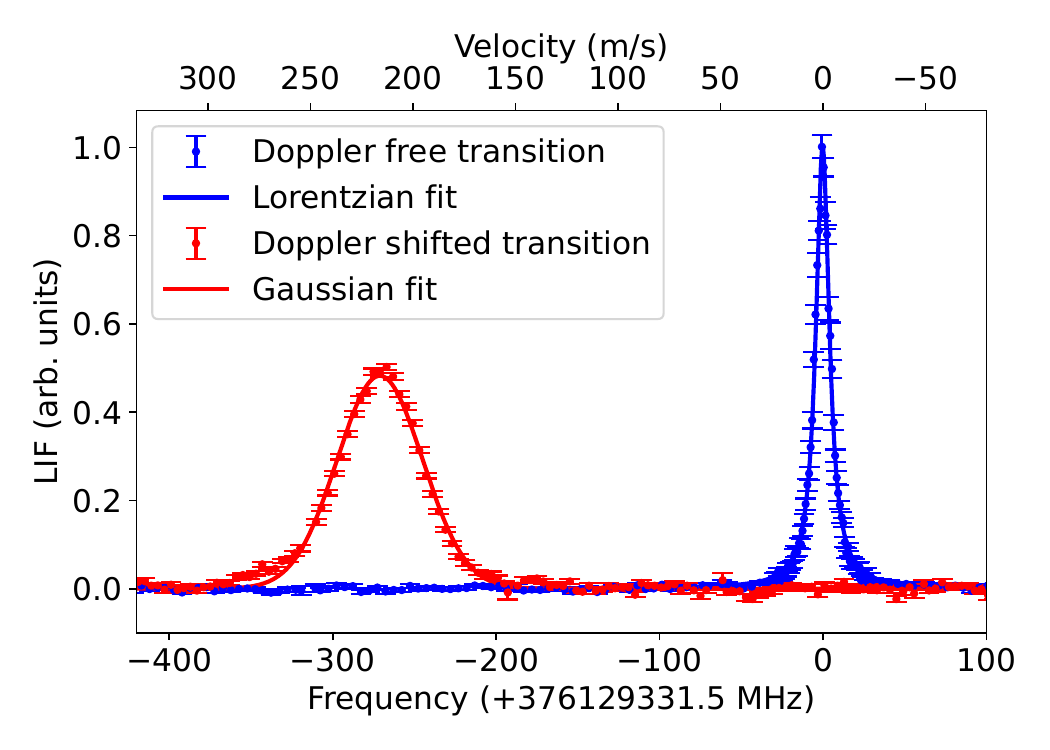}
    \caption{Doppler free and Doppler shifted spectrum of the $A-D$ transition (vertical axis not to scale). Fluorescence at 413\;nm is detected while the frequency of the laser that drives the $A-D$ is scanned and the other laser is locked to the $X-A$ transition frequency. The red data shows the result when the laser beam to drive the $A-D$ transition is counter-propagating with respect to the molecular beam and the laser beam to drive the $X-A$ transition is perpendicular, while the blue data shows the result when both lasers are aligned perpendicularly to the molecular beam. At a certain frequency molecules with a specific longitudinal velocity are Doppler shifted into resonance, as indicated on the top axis. The Doppler-shifted and Doppler-free data are fitted with a Gaussian and a Lorentzian function, respectively. 
    }
    \label{fig:velocity}
\end{figure}

In this section, we will discuss a measurement of the Doppler profile, by using a laser beam counter-propagating with respect to the molecular beam to drive the $A-D$ transition. First, we ensure that the laser beam that drives the $X-A$ transition is aligned to be perfectly perpendicular to the molecular beam by minimizing the frequency difference between the measurements taken with the laser beam from either side. In our velocity measurements, the frequency difference between left and right was typically $\sim$300\;kHz, corresponding to an error in the measured velocity of $\sim$0.1\;m/s. 
Furthermore, we ensure that the laser that drives the $A-D$ transition is perfectly counter-propagating with respect to the molecular beam by aligning it onto the orifice of the source. We estimate the angle between the molecular beam and the laser beam to be below 0.3\;degrees. This corresponds to an error in the measured velocity of below 3$\times10^{-3}$\;m/s. 
Finally, we fix the frequency of the laser that drives the $X-A$ transition to the value listed in Table~\ref{tab:energies} and scan the frequency of the laser that drives $A-D$ transition, while monitoring the fluorescence yield at 413\;nm. 

A typical result of such a measurement is shown by the red data points in Fig.~\ref{fig:velocity}. We scan rapidly (within 4 minutes) over the frequency range and average multiple (20 in this case) scans, to average out source fluctuations caused mainly by variation of the barium yield as a function of the rotation of the ablation target. The error bar shown in the figure represents the standard error of the mean of the signal. The spectrum is fitted by a Gaussian function with a mean detuning of (minus) 271\;MHz and a full width at half maximum (FWHM) of 61\;MHz. From this, we find a mean forward velocity for the molecular beam equal to $\overline{v} = -c( \overline{f} -f_0)/f_0 = 216$\;m/s, with $f_0$ being the $A-D$ Doppler-free resonance, $\overline{f}$ the Doppler shifted resonance frequency and $c$ the speed of light. Similarly, we find that the FWHM velocity spread of the molecular beam is 49\;m/s. 

For reference, we also show a spectrum recorded when both lasers are aligned perpendicular to the molecular beam (blue data) corresponding to the Doppler-free transition. This measurement confirms there are no other features in the spectrum that might complicate the interpretation of the velocity measurement.

It is obvious from the measured average velocity that the molecular beam is strongly boosted during the expansion -- for reference, the mean thermal velocity of BaF molecules and neon atoms at 20\;K is 52\;m/s and 145\;m/s, respectively. This is also obvious from the velocity spread that corresponds to a longitudinal temperature in the moving frame equal to 8\;K, significantly below the temperature of the cell.

\begin{figure}[t]
    \centering
    \includegraphics[width=\columnwidth]{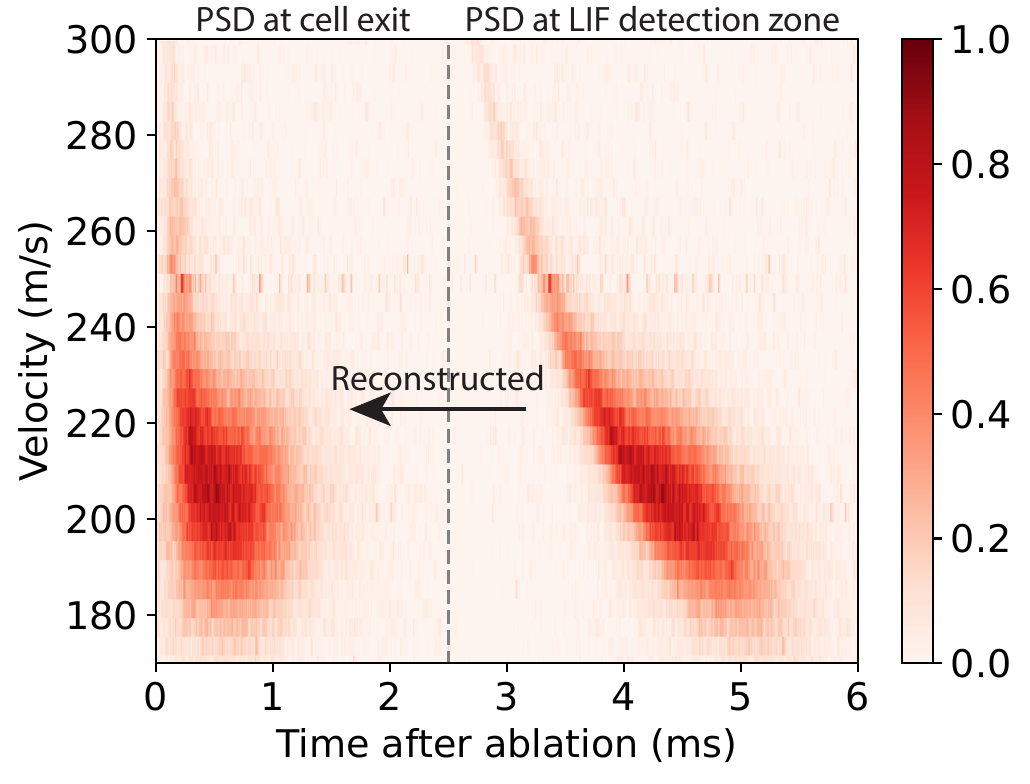}
    \caption{Phase space distribution (PSD) of the molecular beam at two positions along the molecular beam path. On the right-hand side is plotted, the fluorescence signal of the molecules, recorded by the PMT at a distance of 780 mm from the source. On the left-hand side is plotted, the phase space distribution, reconstructed at the exit of the source. The horizontal line observed at 250\;m/s is due to difficulties in determining the frequency of the laser that drives the $A-D$ transition when its beat note with the frequency comb is equal to the repetition rate of the frequency comb. 
    }
    \label{fig:phase-space-reconstructed}
\end{figure}

\section{Measuring the phase-space distribution of a cryogenic buffer-gas beam }\label{sec:demonstration}

Rather than summing the fluorescent signal of the molecules over the pulse duration, such as done in Fig.~\ref{fig:velocity}, we may also plot the velocity as a function of time, as is depicted on the right-hand side of Fig.~\ref{fig:phase-space-reconstructed}. As may be expected, there is a correlation between the time the molecules arrive at the detector and their velocity, i.e., faster molecules arrive earlier and slower molecules arrive later. As the velocity and time-of-flight are measured independently of each other, the phase-space distribution can be reconstructed at any position $z$ along the molecular beam path, and most interestingly, at the source exit by using the expression $t(z=0)=t(z=L_{\mathrm{det}})- L_{\mathrm{det}}/ v$, with $L_{\mathrm{det}}$=780\;mm. This reconstructed phase space density is shown on the left-hand side of Fig.~\ref{fig:phase-space-reconstructed}. As may be observed, the correlations between the arrival time and velocity are much reduced, but are still clearly present. Particularly, it is observed that at the beginning of the pulse, the velocity of the molecules is significantly larger than later in the pulse. This is attributed to the limited heat conduction of the wall of the cell; the temperature of the neon buffer gas is increased by the ablation pulse and it 
takes typically a few hundred microseconds before this heat is transferred to the cell. In a forthcoming paper, we will present a detailed discussion of this process and how it is affected by the source parameters. The time-resolution of the recorded phase-space distribution is limited by the time the molecules take to traverse the laser beam that drives the $X-A$ transition and the $RC$ time of the detection system, which add up to $\sim$13\;$\upmu$s. 
From the sharpest features observed, we deduce an upper limit of the velocity resolution of $\sim$6\;m/s, limited by the natural linewidth of the $X-D$ transition or any remaining power broadening. 
The time-resolution of reconstructed phase-distribution is dominated by the velocity resolution and is about 120\;$\upmu$s.  

\balance

As a check of the validity of our method, we have also reconstructed the phase-space distribution at the longitudinal position where the absorption laser crosses the molecular beam. The red curve in Fig.~\ref{fig:absReconstucted} shows the integrated velocity distribution while the blue curve shows the absorption profile measured simultaneously (averaged over the full duration of the measurements, i.e. over $\sim$10000 shots). The vertical axes have been adjusted such that the amplitudes of the two curves are the same. Although small differences may be observed, the overall agreement is excellent~\footnote[4]{When the neon flow rate is increased, the density in the beam is such that collisions are still taking place after crossing the absorption laser. In this case, the two profiles will be less similar, as is indeed observed in our measurements.}. The main difference is that sharp features are washed out due to the limited velocity resolution of the fluorescence measurements, causing the non-physical result that BaF signal is observed before the ablation laser is fired.

\section{Conclusions}

We have demonstrated a novel method to determine the longitudinal phase-space distribution of a molecular beam via a two-step laser excitation scheme. We use this method to measure the phase-space distribution of a cryogenic buffer gas beam source for barium-fluoride molecules. A strong correlation has been observed between the velocity and time that the molecules leave the source. We obtain a temporal and velocity resolution of 11\;$\upmu$s and 6\;m/s, respectively, limited by the size of the laser beam driving the $X-A$ transition and the spectral resolution of the $A-D$ transition. Our method avoids problems with optical pumping and can also be applied to molecules with hyperfine structure. We plan to use this method under varying conditions to optimise the intensity of slow molecules produced by our beam source.  

\section*{Acknowledgement}
The NL-\emph{e}EDM consortium receives program funding (EEDM-166 and XL21.074) from the Netherlands Organisation for Scientific Research (NWO). We thank Max Beyer for useful discussions on the spectroscopy of barium fluoride and thank Johan Kos, Rob Kortekaas and Leo Huisman for technical assistance to the experiment.




\section*{Conflicts of interest}
There are no conflicts to declare.



\bibliography{references}
\bibliographystyle{rsc} 

\end{document}